\documentclass[trackchanges,twocolumn]{aastex62}

\usepackage{amsfonts,amsbsy}
\usepackage{amssymb,bm,mathrsfs,bbm,amscd}
\usepackage{mathrsfs}
\usepackage{bm} 

\usepackage{CJK,upgreek,fancyhdr}
\usepackage[tbtags]{amsmath}

\usepackage{dcolumn} 
\usepackage{graphicx}
\usepackage{xcolor}
\usepackage{color}

\usepackage{url}
\usepackage{multirow}

\usepackage{ulem}
\usepackage{enumerate}
\usepackage{textcmds}
\usepackage{mathtools}
\usepackage{booktabs}
\usepackage{tabularx}
\usepackage{lipsum}
\usepackage{lastpage}
\usepackage{stfloats}

\usepackage{aas_macros}

\turnoffedit

\renewcommand{\vec}[1]{\boldsymbol{#1}}

\def\kpc{{\rm kpc}}

\newcolumntype{p}{D{,}{\pm}{-1}}

\renewcommand{\figurename}{Fig.}
\renewcommand{\tablename}{Tab.}

\renewcommand{\arraystretch}{1.5}

\revised{1.0 2019.10.30}

\begin{document}

\title{Two numerical methods for the 3D anisotropic propagation of Galactic cosmic rays}

\correspondingauthor{Su-jie Lin, Hong-bo Hu, Yi-Qing Guo, Ai-feng Li}
\email{linsj@ihep.ac.cn, huhb@ihep.ac.cn, guoyq@ihep.ac.cn, liaf@sdau.edu.cn}

\author{Wei Liu}
\affiliation{Key Laboratory of Particle Astrophysics,
Institute of High Energy Physics, Chinese Academy of Sciences, Beijing 100049, China
}

\author{Su-jie Lin}
\affiliation{Key Laboratory of Particle Astrophysics,
Institute of High Energy Physics, Chinese Academy of Sciences, Beijing 100049, China
}

\author{Hong-bo Hu}
\affiliation{Key Laboratory of Particle Astrophysics,
Institute of High Energy Physics, Chinese Academy of Sciences, Beijing 100049, China
}
\affiliation{University of Chinese Academy of Sciences, Beijing 100049, China
}

\author{Yi-qing Guo}
\affiliation{Key Laboratory of Particle Astrophysics,
Institute of High Energy Physics, Chinese Academy of Sciences, Beijing 100049, China
}

\author{Ai-feng Li}
\affiliation{College of Information Science and Engineering, Shandong Agricultural University, Taian 271018, China
}

\begin{abstract}
Conventional cosmic-ray propagation models usually assume an isotropic diffusion coefficient to account for the random deflection of cosmic rays by the turbulent interstellar magnetic field.
Such a picture is very successful in explaining many observational phenomena related to the propagation of Galactic cosmic rays, such as broken power-law energy spectra, secondary-to-primary ratios, etc.
However, the isotropic diffusion presupposition is facing severe challenges from recent observations.
In particular, such observations on the large-scale anisotropy of TeV cosmic rays show that the dipole direction differs from the prediction of the conventional model.
One possible reason is that the large-scale regular magnetic field, which leads to an anisotropic diffusion of cosmic rays, has not  been included in the model provided by the public numerical packages.
In this work, we propose two numerical schemes to solve the $3$-dimensional anisotropic transport equation: the pseudo source method and Hundsdorfer-Verwer scheme. Both methods are verified by reproducing the measured B/C and proton spectrum and the radial variation of spectral index expected by former 2D simulation.
As a demonstration of the prediction capability, dipole anisotropy is also calculated by a toy simulation with a rough megnetic field.
\end{abstract}


\section{Introduction}
\label{sec:intro}

It has long been recognized that after escaping from the acceleration sites, Galactic cosmic rays (GCRs) undergo frequent scatterings from random magnetic turbulence, which could be described by a diffusion process. In the conventional model, the diffusion process is supposed to be uniform and isotropic, and the diffusion coefficient is only rigidity-dependent; namely, $D({\cal R}) \propto {\cal R}^\delta$, with $\delta \sim 0.3 - 0.6$ inferred from the boron-to-carbon ratio \citep{2017PhRvD..95h3007Y}. The propagated CR spectrum falls off as $\phi \propto {\cal R}^{-\nu -\delta}$, where $\nu$ is an injection power index. This CR transport picture has successfully reproduced many basic observational features such as the power-law form of the energy spectra, the secondary-to-primary ratio, and the large-scale distribution of diffuse radio/gamma-ray radiation.



In the past decades, the high-precision and two dimensional observations of CR anisotropy \citep{2005ApJ...626L..29A, 2006Sci...314..439A, 2007PhRvD..75f2003G, 2008PhRvL.101v1101A, 2009ApJ...698.2121A, 2010ApJ...711..119A, 2010ApJ...718L.194A, 2011ApJ...740...16A, 2012ApJ...746...33A, 2013ApJ...765...55A, 2013PhRvD..88h2001B, 2014ApJ...796..108A, 2015ApJ...809...90B, 2016ApJ...826..220A, 2017ApJ...836..153A} disfavored such a simple picture. Neither the amplitude nor the phase could be reproduced by the conventional model \citep{2012JCAP...01..011B, 2017PhRvD..96b3006L}.

Based on the conventional propagation model in which the CR particles are assumed to isotropically diffuse, various improvements have been explored to explain the anisotropy problem, including ensemble fluctuations of CR sources,  nearby source \citep{2006APh....25..183E, 2012JCAP...01..011B, 2013APh....50...33S, 2017PhRvD..96b3006L}, spatially dependent diffusion process \citep{2012PhRvL.108u1102E, 2012ApJ...752L..13T, 2016ApJ...819...54G} etc. Among these models, only the models that involve the nearby source can simultaneously explain both the amplitude and phase of the observed anisotropy well \citep{2019JCAP...10..010L, 2019JCAP...12..007Q}. In this kind of explanation, the expected phase is sensitive to the direction of the nearby source.

As an alternative to these models, the local regular magnetic field responsible for the anisotropic diffusion has been introduced to understand the phase of CR anisotropy. Through observing the ribbon of enhanced energetic neutral atom (ENA) flux, the Interstellar Boundary Explorer (IBEX) experiment \citep{2013ApJ...776...30F} unveiled that within $20$ pc, the local regular magnetic field is along $(l,b) = (210.5^\circ,-57.1^\circ)$, which is coincident with the observed phase of the large-scale anisotropy at TeV energies. 
Several studies \citep{2014Sci...343..988S, 2015PhRvL.114b1101M, 2016PhRvL.117o1103A} propose that the local regular magnetic field may have a projection effect on CR streaming.
This projection effect on CR streaming, together with the contribution from nearby SNRs, may phenomenologically describe both the amplitude and the phase of anisotropy. To study these kinds of models accurately and coherently with constraints from all other observational results, it is necessary to implement the anisotropic propagation in public numerical packages such as GALPROP and DRAGON.

However, it is more complicated to tackle the cross-derivative terms in the anisotropic propagation equation with the traditional finite difference scheme. One popular alternative is the stochastic differential equations (SDEs) approach~\citep{2017SSRv..212..151S, 2018MNRAS.477.1258A}. Based on the equivalence between Fokker-Planck type equations and SDEs, the propagation equation could be recast in a set of SDEs, which is easier to deal with numerically than the original partial differential equation. This approach has been extensively used in the study of solar modulation for low energy CRs, i.e. transport of CRs within the interplanetary magnetic field \citep{2011ApJ...735...83S, 2012ApJ...745..132B, 2013PhRvL.110h1101M, 2016CoPhC.207..386K, 2018ApJ...854...94B}. Nevertheless, compared with the finite difference scheme, it is computationally expensive and time-consuming.


Recently, a 2D numerical implementation of solving the anisotropic diffusion equation based on the finite difference method has been investigated by \cite{2017JCAP...02..015E}, but the corresponding numerical package has not been published yet.
In this work, we propose two recipes to solve a full 3D anisotropic propagation equation: the pseudo source method and the Hundsdorfer-Verwer finite difference scheme.
Demonstrated by a toy magnetic field model, we verified the validity of both methods.
Then we applied our numerical methods to the cosmic-ray phenomena preliminarily.
Both the proton energy spectrum and B/C ratio are calculated under a realistic magnetic field model to compare with observations.
It is found that the propagated spectrum at the solar system is dominated by the diffusion process perpendicular to the regular magnetic field.
And to reproduce the B/C ratio, the propagation parameter $D_{0\parallel}$ has a significant inverse relationship with $\varepsilon = D_{0\perp}/D_{0\parallel}$.
Furthermore, to demonstrate the effect of the regular magnetic field on CR propagation, we also performed a toy simulation, in which the large-scale distribution of CR anisotropy was computed with the Hundsdorfer-Verwer finite difference method according to the fitted transport parameters.
The dipole anisotropy turned out to point toward the direction of regular magnetic field rather than the Galactic center (GC) predicted by the conventional propagation model.

    

The rest of the paper is organized as follows: In Sec. 2, the anisotropic diffusion equation and the two numerical solutions are introduced. The realistic regular magnetic field configuration is also described in detail. Sec. 3 presents the calculation results using two magnetic field models and compares the proton spectrum, B/C ratio with relevant observations. And the dipole CR anisotropies under large-scale regular magnetic field are computed. Sec. 4 is reserved for the conclusion.

\section{Numerical Methods}
\label{sec:method}

\subsection{Anisotropic diffusion}
\label{subsec:3daniso}


A comprehensive introduction to GCR production and propagation can be found in many monographs and review papers, such as \cite{1990acr..book.....B, 2002astro.ph.12111M, 2002cra..book.....S, 2007ARNPS..57..285S}. Here we give a brief review on GCR propagation. The whole diffusive region of GCRs, also called a magnetic halo, is usually approximated as a cylinder with its radial boundary equal to the 
Galactic radius, i.e. $R = 20$ kpc. Its half thickness $L$, which characterizes the vertical stretch of the interstellar magnetic field, is unknown, but is determined by fitting the B/C ratio along with diffusion coefficient. Both CR sources and the interstellar medium (ISM) are chiefly distributed in the Galactic disk with average thickness $z_s$ of roughly $200$ pc. The transport process of CRs in the magnetic halo is described by the following convection-diffusion equation
\begin{align}
 \nonumber  \dfrac{\partial \psi( \vec{r},p,t)}{\partial t} &=
   q(\vec{r},p,t) + \nabla \cdot (D \nabla \psi)  - \nabla \cdot (\vec{V}_{c}\psi)\\
\nonumber & +   \dfrac{\partial}{ \partial p} p^{2}D_{pp} \dfrac{\partial}{\partial p}\frac{1}{p^{2}}\psi
    - \dfrac{\partial}{\partial p} \left[\dot{p}\psi - \dfrac{p}{3}(\nabla \cdot V_{c}\psi) \right] \\
    & -  \dfrac{\psi}{\tau_{f}} - \dfrac{\psi}{\tau_{r}} ~.
\label{eq:con_diffu}
\end{align}
Here, $\psi(\vec{r},p,t)$ is the CR density per unit energy at time $t$. The terms containing $\vec{V}_c$, $D_{pp}$, $\dot{p}$, $\tau_f$, and $\tau_r$ describe the convection, diffusive re-acceleration, energy-loss, fragmentation, and radiative decay effects, respectively. $D$ is the so-called spatial diffusion coefficient, which is in general expressed as a rank-two symmetric tensor. Thus, the diffusion term is written as
\begin{equation}
\nabla \cdot (D \nabla  \psi ) = \frac{\partial}{\partial x_i}\left(D_{ij}\,\frac{\partial\, \psi}{\partial x_j}\right) ~,
\label{eq:diffu_tensor}
\end{equation}
For example, for the two optional coordinate systems in the GALPROP and DRAGON packages, ($x_1, x_2$) denotes ($r$, $z$) in the cylindrical system, while  ($x_1, x_2, x_3$) denotes ($x, y, z$) in the Cartesian coordinate system. When $D_{ij} = D \delta_{ij}$, the diffusion is isotropic; otherwise, it is anisotropic. The off-diagonal elements of $D$ correspond to cross-derivative terms in  Eq.~\ref{eq:diffu_tensor}, which are not included in the Crank-Nicolson scheme. Therefore, the GALPROP that is based on the Crank-Nicolson scheme can only solve the anisotropic diffusion equation in cases where all the off-diagonal elements of $D$ are zero. Besides these cases, the numerical solution for general 2D cases has also been developed by DRAGON~\citep{2017JCAP...10..019C}.


The ordered magnetic field plays an important role in the anisotropic diffusion. In an environment filled with both a regular and irregular magnetic field component, the trajectory of a charged particle can be precisely traced via numerical calculation \citep{2012JCAP...07..031G}. According to these studies, an anisotropic diffusion coefficient becomes necessary when the regular magnetic field strength is comparable to or beyond that of the turbulent one.  The diffusion coefficient tensor can be diagonalized at every point by properly choosing the local coordinate, with one axis parallel and the other two perpendicular to the direction of the local regular magnetic field. In this local coordinate system, the diffusion tensor looks like
\begin{equation}
D_{ij} = \renewcommand{\arraystretch}{0.7}
\begin{pmatrix}
D_\parallel & 0 & 0 \\
0 & D_\perp & 0 \\
0 & 0 & D_\perp
\end{pmatrix} ~.
\end{equation}
Here $D_{\parallel}$ and $D_{\perp}$ are the diffusion coefficients aligned parallel and perpendicular to the ordered magnetic field, respectively. When transforming the above local coordinate system defined by the regular magnetic field to a global one, $D_{ij}$ turns to be \citep{1999ApJ...520..204G}
\begin{equation}
D_{ij}\,\equiv\,D_\perp\delta_{ij}\,+\,\big(D_\|-D_\perp\big)b_ib_j ~,
\label{eq:D_ij_1}
\end{equation}
where $b_i = \dfrac{B_i}{|\vec{B}|}$ is the $i$-th component of the unit vector of the ordered magnetic field $\vec{B}$ in the chosen coordinate system. 

In the Galaxy, the turbulent fluctuations of the magnetic field ($\vec{B}$) are regarded to be injected at a scale of 100 pc with a strength $\delta B/B_0\sim1$. At the energy region $\sim$ GeV, the mean free path of the CR particles is $\sim1$ pc. The turbulent power $\delta B$ at this scale would be largely suppressed as a result of the turbulent cascade, i.e. $\delta B/B_0 \ll 1$. This small turbulence could be dealt with by the quasi-linear theory (QLT) \citep{1966ApJ...146..480J,jokipii1968random}, in which the ratio between the perpendicular and parallel diffusion coefficients is expected to be
\begin{equation}
\dfrac{D_\perp}{D_\parallel} \sim \mathcal{F}(k) \sim \dfrac{\delta B_k^2}{B_0^2} \ll 1 ~,
\end{equation}
where $\mathcal{F}(k)$ is the normalized power of the turbulent modes with wave number $k$. For higher energies, this ratio needed to be calculated with the realistic simulation. In a recent numerical simulation, it was shown that this ratio is always smaller than $\mathcal{O}(0.1)$ for $\delta B / B_0 \le 1$~\citep{2007JCAP...06..027D} and the perpendicular diffusion always has a steeper rigidity dependence than the parallel component \citep{2007JCAP...06..027D, 2002PhRvD..65b3002C, 2016MNRAS.457.3975S}.

In this work, we simply adopt two different power-law dependents $D_\parallel$ and $D_\perp$ following \cite{2017JCAP...10..019C},
\begin{align}
D_\parallel &= D_{0\parallel} \left(\frac{\cal R}{{\cal R}_0} \right)^{\delta_\|} ~, \\
D_\perp &=\,D_{0\perp} \left(\frac{\cal R}{{\cal R}_0} \right)^{\delta_\perp} \equiv \varepsilon D_{0\parallel} \left(\frac{\cal R}{{\cal R}_0} \right)^{\delta_\perp} ~,
\label{eq:DparaDperp}
\end{align}
in which $\varepsilon = \dfrac{D_{0\perp}}{D_{0\parallel} }$ is the ratio between perpendicular and parallel diffusion coefficient at the reference rigidity ${\cal R}_{0}$. As discussed above, in order to make a simulation physically meaningful,  the parameters $\delta_\parallel$, $\delta_\perp$, $R_0$, and $\varepsilon$ need to be carefully chosen to keep $D_\perp\lesssim D_\parallel$ within the concerned energy region.  In order to show how the two diffusion coefficients affected the result, in the calculation of this work, we would choose several sets of benchmark parameters that may be physically non-realistic.

\subsection{Solutions to cross-derivative diffusion terms}

As shown above, the off-diagonal elements of the diffusion tensor $D$ are inevitable in general cases.  Therefore, a new algorithm has to be developed to take care of the corresponding cross-derivative terms.

In this work, we try to deal with a full 3D anisotropic propagation equation with two different methods. 
One is a pseudo source method which attributes the off-diagonal terms to the source term, and the other one is an implicit calculation based on the Hundsdorfer-Verwer scheme~\citep{HUNDSDORFER2002213}.

\subsubsection{Pseudo source method}

In the pseudo source method, the cross-derivative terms are combined with the source term, based on the numerical package GALPROP.
In other words, the first two terms of the rhs. of equation \ref{eq:con_diffu} become
\begin{align}
\nonumber & q(\vec{r},p,t) + \nabla \cdot (D \nabla \psi) \\
\nonumber & = \left[ q(\vec{r},p,t) + \sum_{i\neq j} \dfrac{\partial}{\partial x_i}  \left(D_{ij} \dfrac{\partial \psi}{\partial x_j}  \right)  \right]  + \dfrac{\partial}{\partial x_i} \left(D_{ii} \dfrac{\partial \psi}{\partial x_i} \right)  \\
\nonumber&= \left[ q(\vec{r},p,t) + q_{\rm pseudo} (\vec{r},p,t) \right]  + \dfrac{\partial}{\partial x_i} \left(D_{ii} \dfrac{\partial \psi}{\partial x_i} \right) \\
&= q^\prime(\vec{r},p,t) + \dfrac{\partial}{\partial x_i} \left(D_{ii} \dfrac{\partial \psi}{\partial x_i} \right)
\label{eq:new_term}
\end{align}
where $q_{\rm pseudo}$ is called the pseudo source, which represents the cross-derivative terms. $q^\prime$ is the sum of the real and pseudo sources. In this way, the 3D propagation equation with cross-derivative terms returns to the diagonal form, which is perfectly fit for the GALPROP package.

The iteration method is used to solve this equation. In the first step, no pseudo source is assumed. We just solve a diffusion equation by neglecting cross-derivative terms and obtaining a first-order CR distribution $\psi^{(1)}$. After the first iteration, the pseudo source $q_{\rm pseudo}$ can be constructed according to the definition in Eq. (\ref{eq:new_term}), by using the CR distribution obtained from the last iteration. After solving this equation, a second-order distribution $\psi^{(2)}$ is obtained. By repeating this process until the solution converges, the final distribution is obtained. 

\subsubsection{Hundsdorfer-Verwer scheme}
Another method to deal with the off-diagonal terms is to replace the Crank-Nicolson scheme used in GALPROP with a finite difference scheme which could handle the cross-derivative terms. Here we adopt the Hundsdorfer-Verwer scheme whose calculation error and convergence is well studied in  Ref.~\citep{HUNDSDORFER2002213}. 

A general $s$-dimensional partial differential equation
\begin{equation}
\frac{\mathrm{d}\psi}{\mathrm{d}t} = F(t, \psi(t))
\end{equation}
could be solved with the operator splitting method. In this method, we decompose the function $F$ into a sum of simpler components
\begin{equation}
F(t,\psi) = F_0(t,\psi) + F_1(t,\psi) + \dots + F_s(t,\psi),
\end{equation}
where the component $F_i$ for $i=1,2,\dots, s$ is supposed to contain all the terms of the pure derivative operators like $\partial_{x_i},\partial_{x_i}^2\dots$\,. Then, we can implement the iteration $\psi_n\rightarrow\psi_{n+1}$ with a simple Douglas scheme~\citep{douglas1962alternating},
\begin{equation}
\left\{\begin{array}{lr}
u_0 = \psi_n + \tau F(t_n,\psi_n), & \\
u_j  = u_{j-1} + \theta\tau\left(F_j(t_{n+1}, u_j) - F_j(t_n,\psi_n)\right) & (j=1,2,\dots,s),\\
\psi_{n+1}=u_s,
\end{array}\right.
\label{eq:Douglas-scheme}
\end{equation}
where $\tau$ is the time step size and $\theta\in[0.5, 1]$ is an auxiliary parameter to make the iteration implicit. In the case $\theta=\frac{1}{2}$, this iteration is similar to the Crank-Nicolson scheme.

All the terms in $F_i (i=1,2,\dots,s)$ are explicitly calculated in the first line and then implicitly corrected in the second line of Eq.~\ref{eq:Douglas-scheme}. Without cross-derivative terms, $F_0$ is independent of $\psi$ and the iteration is thus totally implicit. However, when the cross-derivative terms are considered, they would be absorbed into $F_0$ as explicit terms. The Ref.~\citep{HUNDSDORFER2002213} has studied the stability of the iteration with such explicit terms and has suggested the Hundsdorfer-Verwer scheme inspired by the Runge-Kutta method 
\begin{equation}
\left\{\begin{array}{lr}
u_0 = \psi_n + \tau F(t_n,\psi_n), & \\
u_j  = u_{j-1} + \theta\tau\left(F_j(t_{n+1}, u_j) - F_j(t_n,\psi_n)\right) & (j=1,2,\dots,s),\\
\psi^*_{n+1}=u_s, & \\
u^*_0 = \psi_n + \frac{1}{2}\tau(F(t_n,\psi_n)+F(t_{n+1},\psi^*_{n+1}),&\\
u^*_j  = u^*_{j-1} + \theta\tau\left(F_j(t_{n+1}, u^*_j) - F_j(t_{n+1},\psi^*_{n+1})\right) & (j=1,2,\dots,s),\\
\psi_{n+1} = u^*_s. & \\
\end{array}\right.
\label{eq:HV-scheme}
\end{equation}
Compared to Eq.~\ref{eq:Douglas-scheme},  this scheme which integrates the explicit part in a second-order manner is found to converge faster.

\section{Verifications and possible application}
\label{sec:res}

\subsection{Verification with a toy magnetic field model}
\label{subs:Toy}
Before applying the realistic magnetic field configuration, we adopted a 2D toy model used by Ref.~\citep{2017JCAP...02..015E} and compared our methods with their 2D anisotropic diffusion calculation. The toy model is described as
\begin{align}
\nonumber B_r &= 0 ~, \\
B_\phi &= B_{0,\phi} \left(1 - \exp \left[-\dfrac{r}{r_0} \right]  \right) \\
\nonumber B_z &= B_{0,z} \exp \left[-\dfrac{r}{r_0} \right] \equiv \varepsilon_B B_{0,\phi} \exp \left[-\dfrac{r}{r_0} \right]
\end{align}
in which $\varepsilon_B = \dfrac{B_{0, z}}{B_{0, \phi}}$. 

\begin{figure*}
\centering
\includegraphics[height=6.cm, angle=0]{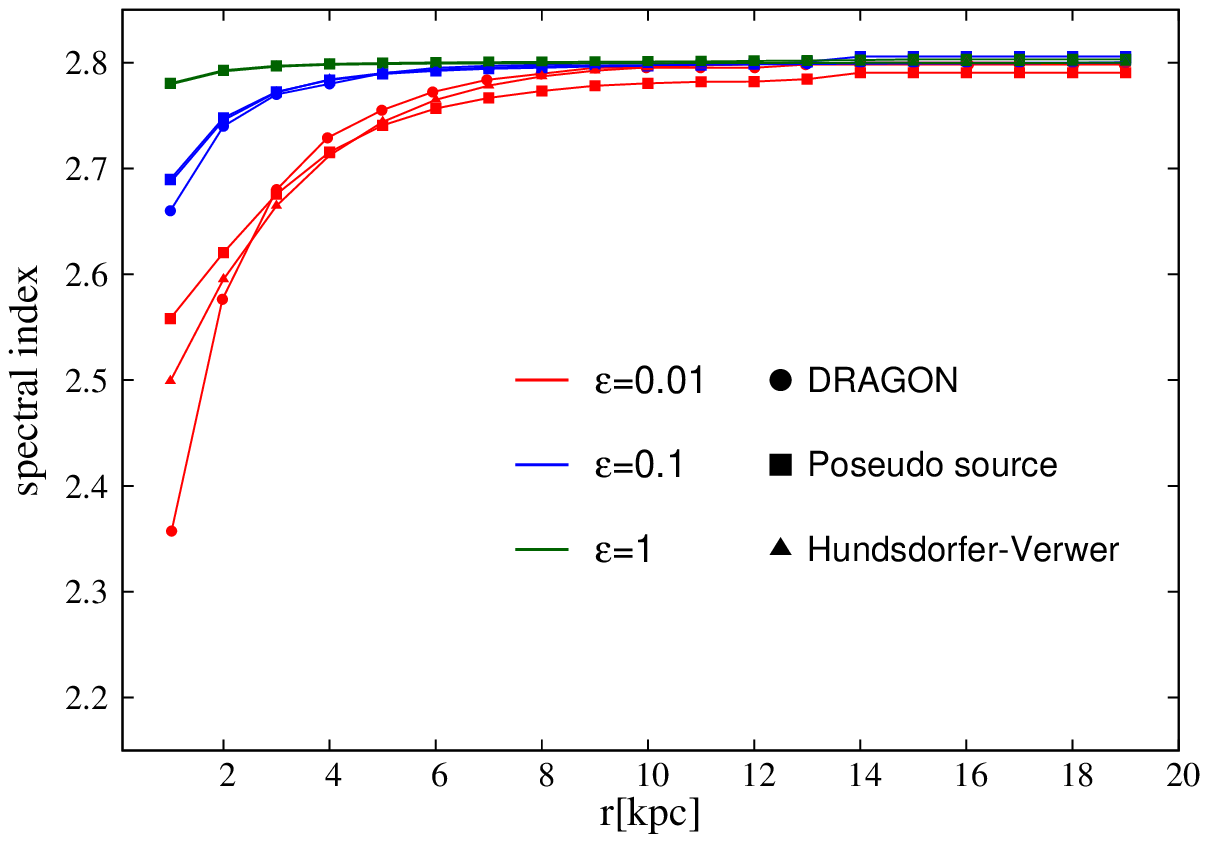}
\includegraphics[height=6.cm, angle=0]{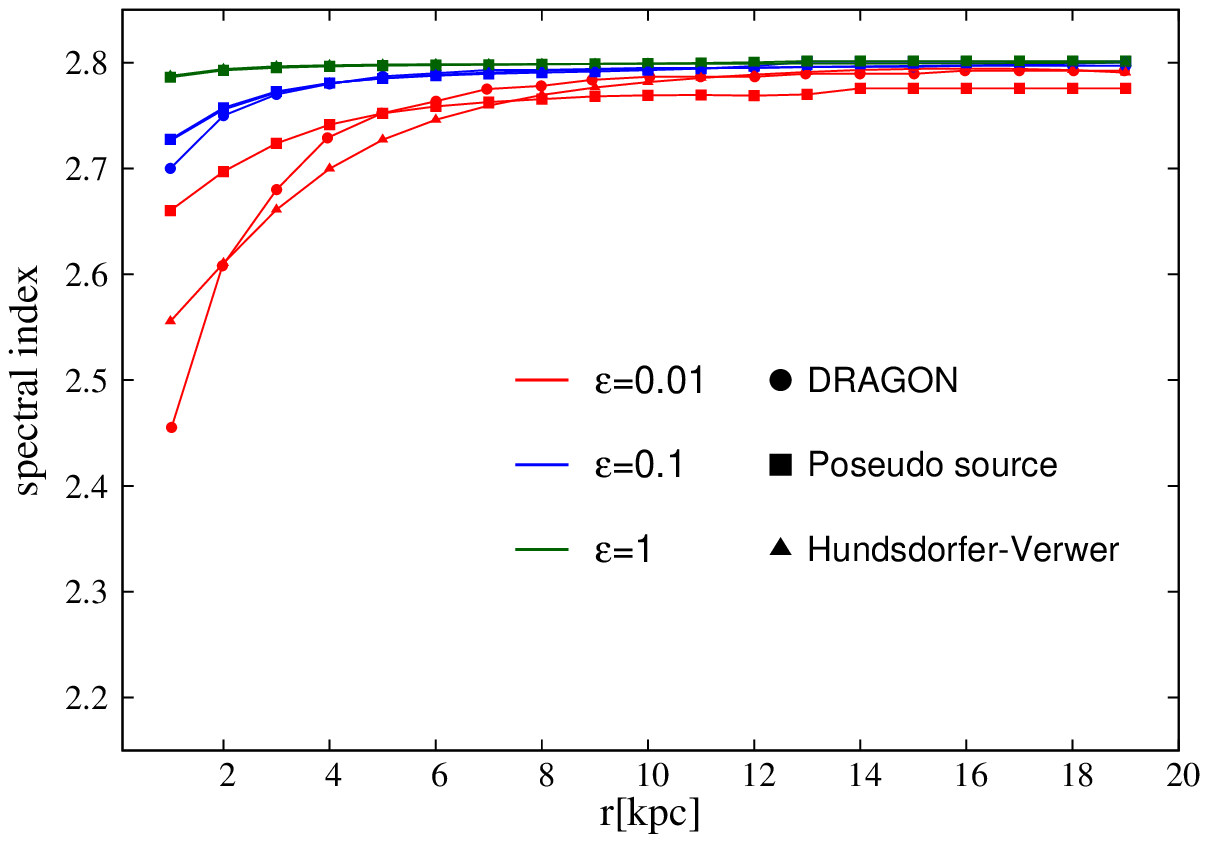} 
\caption{Comparison of radial variation of spectral index. In the left and right figures,  $z_h$ equals to $2$ and $4$ kpc respectively. Here $\delta_\parallel = 0.1$, $\delta_\perp = 0.6$, $R_0 = 3.5$ kpc, and $\varepsilon_B = 0.2$. The colors green, blue, and red indicate the cases with $\varepsilon=1,0.1$, and $0.01$, respectively, while the point shapes circle, square, and triangle indicate the results from DRAGON 2~\citep{2017JCAP...10..019C}, pseudo source method, and Hundsdorfer-Verwer scheme, respectively. 
}
\label{fig:spec_index}
\end{figure*}

This simple toy model could roughly describe the Galactic magnetic field. Close to the GC, $B_z$ is the major component of magnetic field, but $B_\phi$ dominates over $B_z$ far from the GC. Therefore with increasing distance, the CR diffusion gradually changes from parallel diffusion dominated (near the GC) to perpendicular diffusion dominated (at large distances). So, a radial variation of the spectral index of the propagated spectrum is expected if $\delta_\parallel \neq \delta_\perp$, i.e.
\begin{equation*}
\psi({\cal R}) = \left\{ 
\begin{array}{cc}
{\cal R}^{-(\alpha +\delta_\parallel)} ~~ & r \ll r_0 \\
{\cal R}^{-(\alpha +\delta_\perp)}  ~~ & r \gg r_0
\end{array}
\right.,
\end{equation*}
where the the injection spectrum is assumed to be a power-law $\mathcal{R}^{-\alpha}$.
In Fig. \ref{fig:spec_index}, we show the radial variation of the proton's spectral index computed by our methods in comparison with the DRAGON 2 package~\citep{2017JCAP...10..019C}. This result is derived with parameters $\delta_\parallel = 0.1$, $\delta_\perp = 0.6$, $R_0 = 3.5$ kpc, and $\varepsilon_B = 0.2$ for $\varepsilon=1$, $0.1$, and $0.01$, respectively. It can be seen that due to the anisotropic diffusion, the power index of propagated spectrum gradually becomes hard when approaching to the GC. Except for the cases with $\varepsilon=0.01$ in which the difference between 2D and 3D models is obvious, our results are consistent with DRAGON 2.


To compare with the observational data, a realistic distribution of the magnetic field must be used.

\subsection{Regular large scale structure of Galactic magnetic field}
\label{subs:B_field}
Abundant measurements indicate that the Galactic magnetic field has a large-scale regular field, which has a magnitude comparable to the turbulent component \citep{2001RvMP...73.1031F}. Observations of other spiral galaxies also arrive at similar conclusions \citep{2011MNRAS.412.2396F, 2012SSRv..166..215B}. As for our Galaxy, the regular magnetic field contains three components: 1) the disk components are along spiral arms, and are counter-clockwise in all inner arms when viewed from the northern galactic pole. Evidence indicates that in the  inter-arm region, the magnetic field is clockwise in direction. 2) Antisymmetric azimuthal halo or toroidal magnetic field above and below the galactic plane. 3) Poloidal or X-shaped halo magnetic fields cross over the galactic plane. It is nearly perpendicular to the galactic plane when close to the GC, but has a tilted angle further from the GC. In this work, a realistic magnetic field model is adopted according to Ref.~\citep{2017JCAP...10..019C}. The detailed form of magnetic field is described as follows.

The disk component is taken as azimuthally symmetric, which is a function of $r$ and $z$, i.e.
\begin{equation}
B_{\phi}^{\rm disk} \, (r,z) \,= \, \left\{
 \begin{array}{cc}
 B_{0}^{\rm d} \, \exp \left(-\dfrac{\lvert z \lvert}{z_0} \right) \,\, ~, &  r \leqslant R_{c}^{\rm d} \\
  \\
 B_{0}^{\rm d} \, \exp \left(-\dfrac{(r - r_{\odot})}{R_0} -\dfrac{\lvert z \lvert}{z_0} \right) \,\, ~,& r > R_{c}^{\rm d}
 \end{array}
 \right.\,,
 \label{eq:Bdisk}
\end{equation}

The azimuthal halo component is parameterized as
\begin{equation}
B_{\phi}^{\rm halo} \, (r,z) \,= \, B_{0}^{\rm h} \left[ 1 + \left( \frac{\lvert z \lvert  - z_0^{\rm h}}{z_1^{\rm h} } \right) \right]^{-1} \,\frac{r}{R_0^{\rm h} } \, \exp\left(1 - \frac{r}{R_0^{\rm h} }\right) ~.
\label{eq:Bhalo}
\end{equation}
The direction of the magnetic field in the northern hemisphere is counter-clockwise, while it is reversed in the southern part.

The halo poloidal magnetic field adopted here is the parametrized as, 
\begin{equation}\label{eq:BXX_Bz}
B^{\rm pol} \, (r,z) \,= \, B_{\rm X}(r,z) \exp \left( -\dfrac{R^{\rm p}}{R^{\rm X}} \right) \cos\big[\Theta_{\rm X}(r,z)\big]\,,
\end{equation}
\begin{equation}\label{eq:BXX_BR}
B_r^{\rm pol} \, (r,z) \,= \, B_{\rm X}(r,z) \exp \left( -\dfrac{R^{\rm p}}{R^{\rm X}} \right) \sin\big[\Theta_{\rm X}(r,z)\big]\,,
\end{equation}
with $B_{\rm X}$, $\Theta_{\rm X}$ and $R^{\rm p}$ defined as
\begin{equation}
B_{\rm X}(R,z) \,= \, \left\{
 \begin{array}{cc}
  B_0^{\rm X} \, \left(\dfrac{R_p}{r}\right)^2  \,\,~, &  r \leqslant R_c^{\rm X} \\
  \\
  B_0^{\rm X} \, \left(\dfrac{R_p}{r}\right) \,\,~, &  r > R_c^{\rm X}
 \end{array}
 \right.\,,
\label{eq:bX}
\end{equation}

\begin{equation}
\Theta_{\rm X}(r,z) \,= \, \left\{
 \begin{array}{cc}
  \tan^{-1}\left(\dfrac{|z|}{r-R^{\rm p} }\right) \,\,~, &  r \leqslant R_c^{\rm X} \\
  \\
  \Theta_0^{\rm X} \,\, ~, &  r > R_c^{\rm X}
 \end{array}
 \right.\,,
\label{eq:ThetaX}
\end{equation}
and
\begin{equation}
R^{\rm p}\,= \, \left\{
 \begin{array}{cc}
   \dfrac{r R_c^{\rm X}}{R_c^{\rm X}+\dfrac{|z|}{\tan\Theta_0^{\rm X}} }\,\, ~, & r \leqslant R_c^{\rm X} \\
  \\
  r-\dfrac{|z|}{\tan\Theta_0^{\rm X}}\,\, ~,&  r > R_c^{\rm X}
 \end{array}
 \right.\,,
\label{eq:Rp}
\end{equation}
All of the parameters are listed in the table \ref{tab:Bpara}. 

\begin{table*}[t!]
\begin{center}
\begin{tabular}{ccccccccc}
\toprule[1.5pt]
\hline
Disk & $B_0^{\rm d}$ [$\mu$G]  & $z_0$ [kpc] & $R_c^{\rm d}$ [kpc] \\
        & $2.0$  & $1.0$ & $5.0$  \\
\hline
Halo & $B_0^{\rm h}$ [$\mu$G] & $z_0^{\rm h}$ [kpc]  & $z_{1}^{\rm h}$ [kpc] & $R_0^{\rm h}$ [kpc] \\
         & $4$ & $1.3$ & $0.25$ & $8$ \\
\hline
Poloidal & $B_0^{\rm X}$ [$\mu$G] & $\Theta_0^{\rm X}$ & $R_c^{\rm X}$ [kpc] & $R^{\rm X}$ [kpc] \\
              & $4.6$  & $49^\circ$ & $4.8$  & $2.9$ \\
\bottomrule[1.5pt]
\end{tabular}
\end{center}
\caption{Parameters of Galactic magnetic field model.}
\label{tab:Bpara}
\end{table*}

\subsection{B/C ratio and Proton spectrum}
\label{subs:Real}

To study the effect of 3D anisotropic diffusion with the aforementioned magnetic field model, the proton spectrum and B/C ratio are calculated with a plain diffusion model without including the convection and diffusive re-acceleration terms.

\begin{figure*}
\centering
\includegraphics[height=9.cm, angle=0]{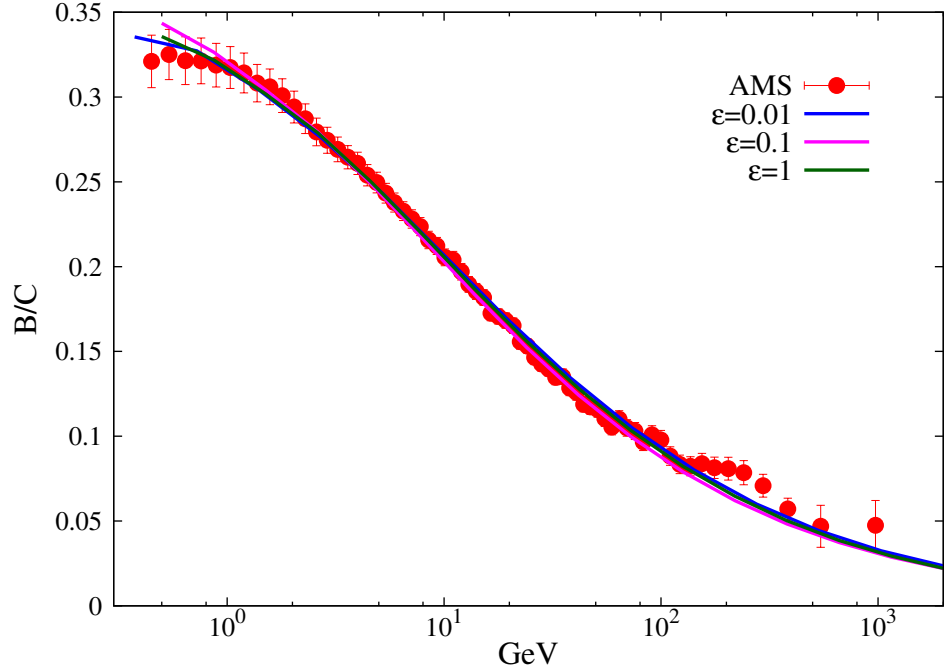}
\caption{Fitting to B/C ratio with different $\varepsilon$ under 3D anisotropic diffusion. The propagation parameters are listed in table \ref{tab:3dpara}. The B/C data points are taken from AMS-02 experiment \citep{2016PhRvL.117w1102A}. 
}
\label{fig:bcratio}
\end{figure*}

Here we show the results calculated with the Hundsdorfer-Verwer scheme; the pseudo source method would lead to similar results.
Fig.~\ref{fig:bcratio} illustrates the fitting of the B/C ratio with different sets of parameters. The propagation and injection parameters are listed in table \ref{tab:3dpara}. As shown in the table, when $\varepsilon$ decreases, both $D_{0\parallel}$ and $\delta_\perp$ gradually increase. In the isotropic diffusion, the ratio of B/C is proportional to $L/D$. In the vicinity of the solar system, the diffusion is dominated by the perpendicular diffusion, $D_\perp$. Thus, $D_\perp$ is determined by the B/C ratio when $L$ is fixed. Then, as the ratio of perpendicular to parallel diffusion $\varepsilon$ drops, $D_{0\parallel}$ has to rise. Under all three cases, the calculated B/C ratio can fit the observations.



\begin{figure*}
\centering
\includegraphics[height=10.cm, angle=0]{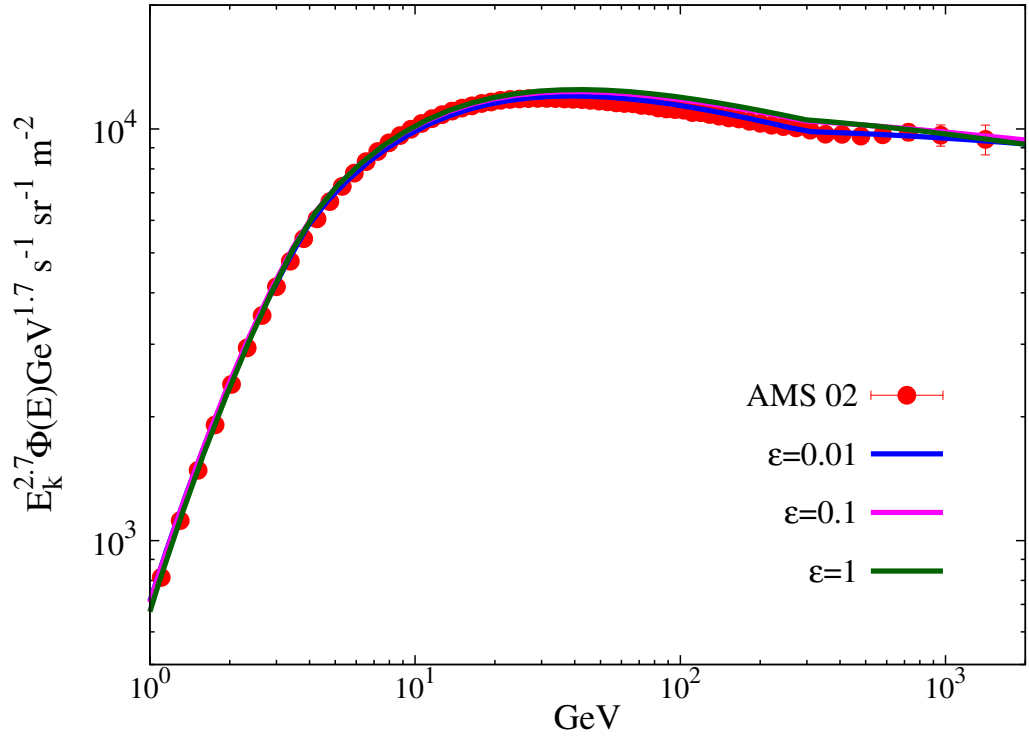} \\
\caption{Calculated proton spectra with different $\varepsilon$ under 3D anisotropic diffusion. The propagation and injection parameters are listed in table \ref{tab:3dpara}. The proton data points are taken from AMS-02 experiment \citep{2015PhRvL.114q1103A}.
}
\label{fig:prot}
\end{figure*}

With the propagated parameters determined by fitting the B/C ratio, the propagated proton spectrum can be obtained. Fig. \ref{fig:prot} shows the propagated proton spectrum with different sets of parameters. Within three kinds of propagated parameters, the calculated spectrum can reproduce the AMS-02 observation. Similar to the B/C ratio fitting, here we also conclude that the propagated spectrum at the solar system is decided by the perpendicular diffusion.


\begin{table*}
\begin{center}
\begin{tabular}{ccccccccccc}
\toprule[1.5pt]
\hline
& $\varepsilon$  ~~& $D_{0\parallel}$ [cm$^2$/s] ~~& ~~~$\delta_{\parallel}$ ~~& ~~~$\delta_{\perp}$ ~~& ~~$\eta$ ~~ & $\nu_1$ & ${\cal R}_1$ [GV] & $\nu_2$ &  ${\cal R}_2$ [GV] & $\nu_3$\\ 
   \hline
Set 1 & 1.0   & $2.7\times 10^{28}$    & 0.3  & 0.64  & $0.5$ & $1.85$  & $5.5$ & $2.34$ & 300 & 2.25 \\
Set 2 & 0.1   & $1.1\times 10^{29}$    & 0.3  & 0.64  & $0.5$ &  $1.9$   & 5.5  &  2.34 & $300$ & $2.25$ \\
Set 3 & 0.01 & $4.0\times 10^{29}$    & 0.3  & 0.64  & $0.5$ &  2.0    & 5.5  & $2.34$ & $300$ & $2.25$  \\
\bottomrule[1.5pt]
\end{tabular}
\end{center}
\caption{Parameters of propagation and injection spectra. The proton flux is normalized to $4.6\times 10^{-9}$ cm$^{-2}$ sr$^{-1}$ s$^{-1}$ MeV$^{-1}$ at $100$ GeV.}
\label{tab:3dpara}
\end{table*}


%

\subsection{Large-scale dipole anisotropy}
\label{subs:aniso}

The anisotropic diffusion model could be applied to explain the dipole anisotropy of CR. Here we show the prospect of this explanation with our current methods. We just take Hundsdorfer-Verwer scheme as an example in this subsection.

The amplitude of the dipole anisotropy is usually defined with observed flux $\bm\phi$ as $|\bm\delta|\equiv(\bm\phi_{\rm max}-\bm\phi_{\rm min})/(\bm\phi_{\rm max}+\bm\phi_{\rm min})$.
In the scheme of diffusive propagation, its vector form can be written as 
\begin{equation}
\bm{\delta} = \dfrac{3D \cdot \nabla \psi}{v \psi} = \dfrac{3}{v \psi} D_{ij}  \dfrac{ \partial \psi}{ \partial x_j} ~.
\end{equation}
Under the scenario of anisotropic diffusion, the CRs tend to propagate along regular magnetic field lines.
This could lead to a larger anisotropy component along the magnetic field.

However, the Galactic magnetic field model described in subsection \ref{subs:B_field} does not resemble the local interstellar magnetic field within $\sim$ 100 pc.
\begin{table}
  \centering
  \begin{tabular}{cccc}
    \hline
    \hline
                      &  $x$  &  $y$  &  $z$\\
    \hline
        Length($\kpc$)  & 32  & 32  &  8\\
        Grid size($\kpc$)  & 0.5  & 0.5  &  0.2\\
    \hline
  \end{tabular}
  \caption{The grid size and total length in $x$, $y$ and $z$ direction in this calcualtion.}
  \label{tab:grid_size}
\end{table}
As mentioned before, the local regular magnetic field within $20$ pc has been measured accurately by IBEX experiment \citep{2009Sci...326..959M}.

However, current computation resources could not implement the computation of anisotropic diffusion in such small regions.
  Limited by the compute resoures, the grid size in the simulation is at least $\sim\mathcal{O}(0.1\kpc)$.
  In Tab.~\ref{tab:grid_size}, we show the propagation halo size and grid size used in this work.
  To manifest the effect of local regular magnetic field, we just modify the magnetic field in the $3\times 3\times 3$ grids around sun.
  According to the grid size adopted here, in another word, the local diffusion region at the solar system is artificially enlarged to a $\mathrm{1.5\,kpc\times1.5\,kpc\times0.6\,kpc}$ cubic volume.
  This calculation is just a rough toy simulation, more effort is needed in the future in order to approach a small enough grid size.

\begin{figure*}
\centering
\includegraphics[height=6.cm, angle=0]{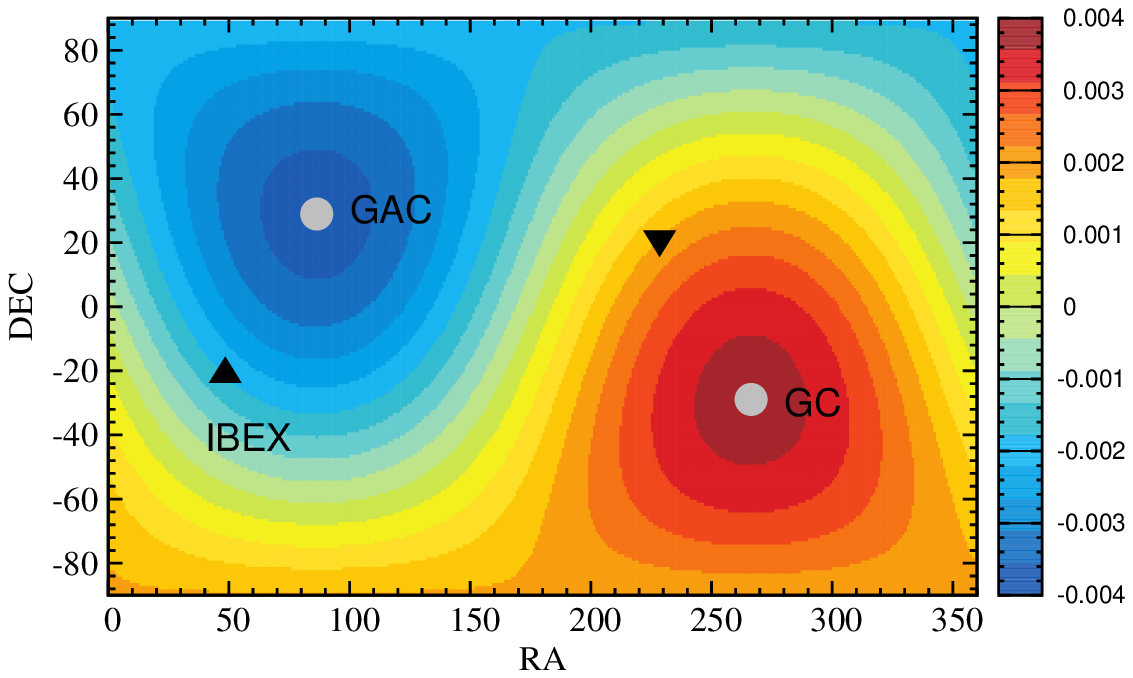} 
\includegraphics[height=6.cm, angle=0]{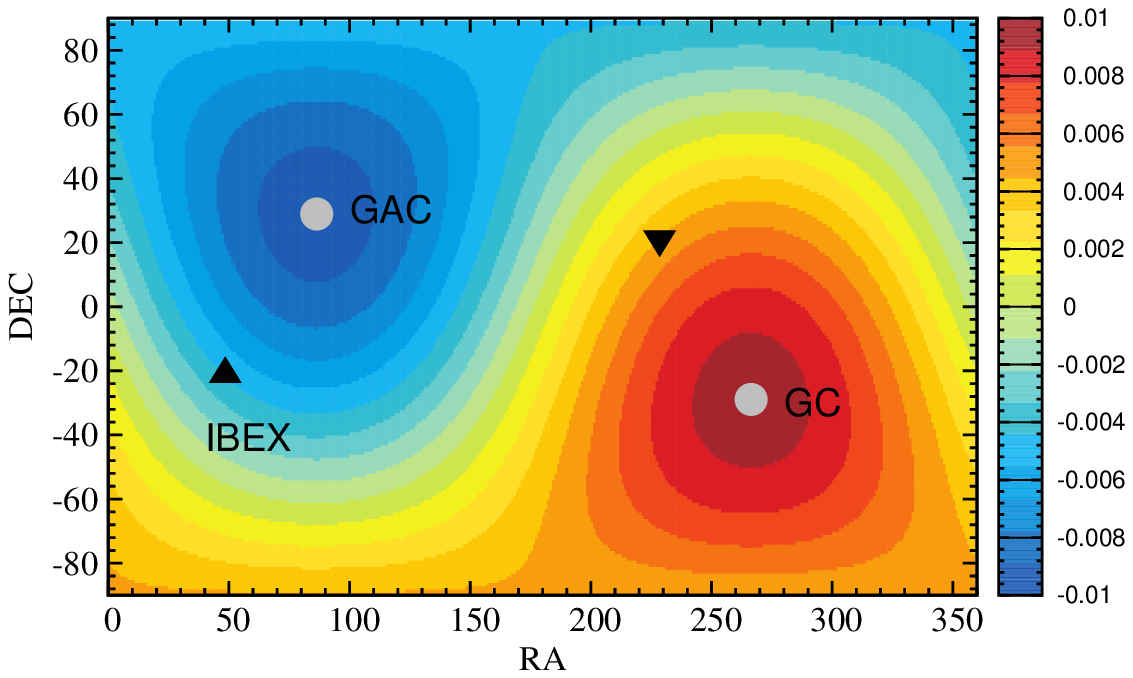} 
\caption{Two-dimensional sky maps of the large-scale dipole anisotropy for the isotropic diffusion scenario at $1$ TeV (left) and $10$ TeV (right) respectively. The gray circles indicate the GC and GAC directions while the black triangle indicate the magnetic field direction measured by IBEX.
}
\label{fig:delta_iso}
\end{figure*}
\begin{figure*}
\centering
\includegraphics[height=6.cm, angle=0]{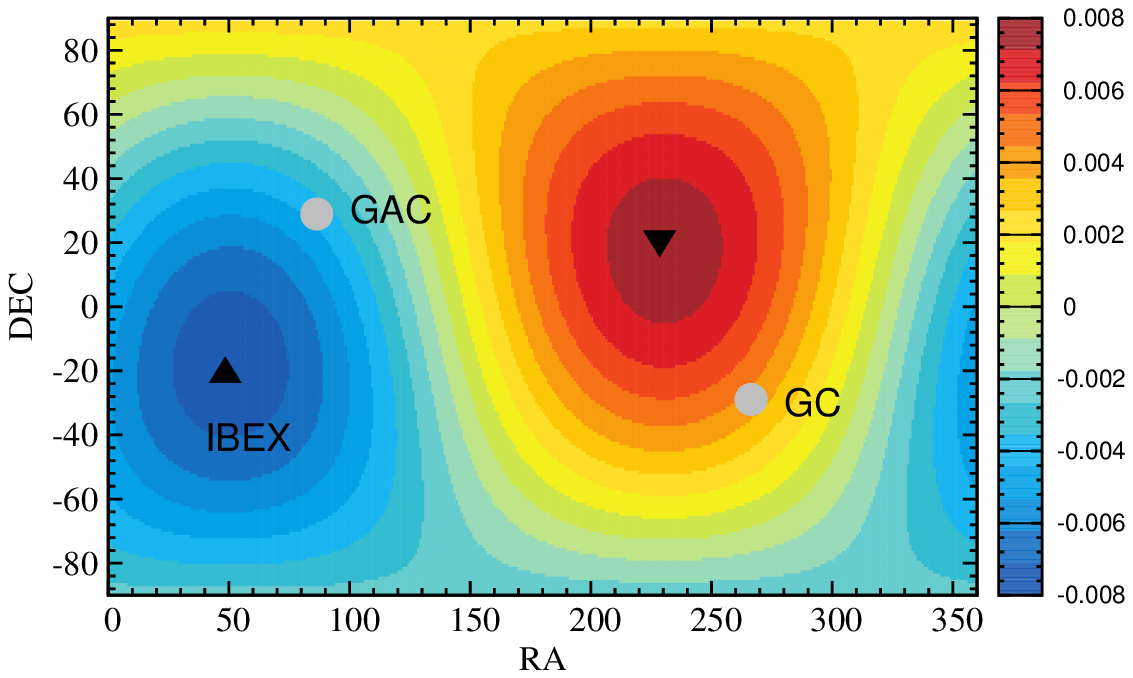}
\includegraphics[height=6.cm, angle=0]{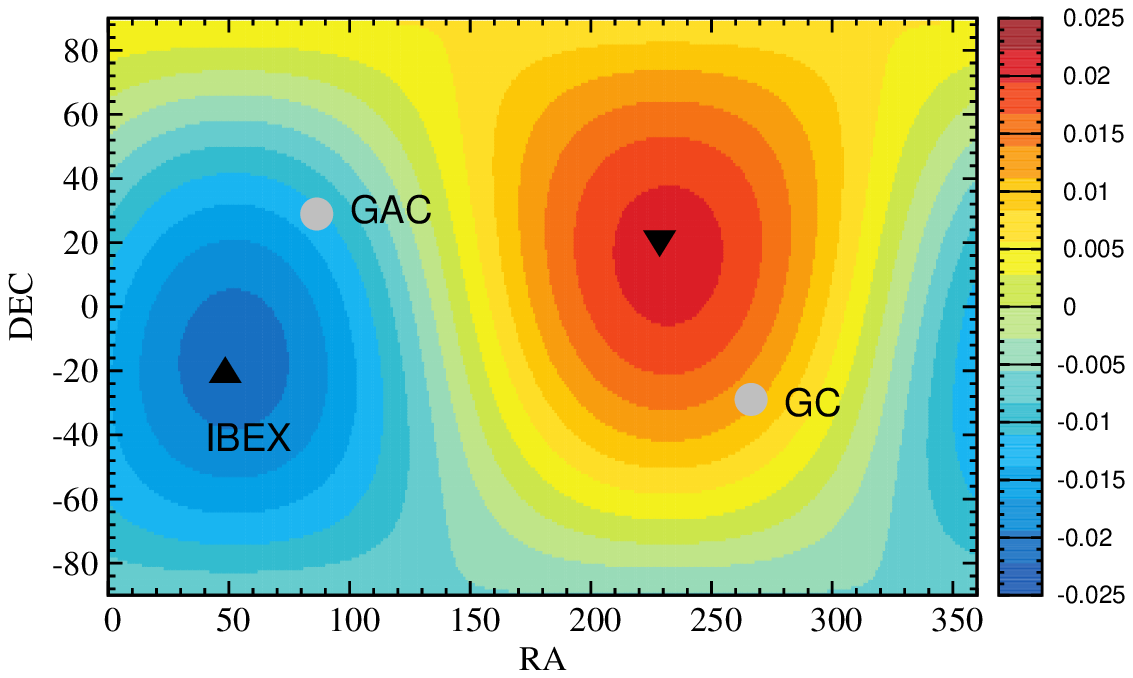}
\caption{
The same as Fig.~\ref{fig:delta_iso} but for the anisotropic diffusion model.
}
\label{fig:delta_ani}
\end{figure*}

Assuming a conventional isotropic diffusion, Fig.~\ref{fig:delta_iso} illustrates the two-dimensional anisotropy skymaps expected at energies $1$ TeV and $10$ TeV, respectively.
Since the source density closer to the GC is higher,  the direction of the excess is always towards the GC.
But it is different for the anisotropic propagation. Using the Set 3 propagation parameters shown in Tab.~\ref{tab:3dpara}, Fig.~\ref{fig:delta_ani} shows the corresponding skymaps resulted from the anisotropic propagation model. In this case, the CR stream is projected along the magnetic field direction, so the excess points along the direction of the local magnetic field. 



\section{Summary and Discussion}
\label{sec:sum}

In this work, a practical pseudo source method and  the Hundsdorfer-Verwer scheme have been proposed to numerically solve the anisotropic diffusion equation. We verified the validity of our methods by comparing with previous works \citep{2017JCAP...10..019C} under a toy magnetic field model and further studying the anisotropic diffusion by using a realistic magnetic field model. The B/C ratio and proton spectrum were calculated, both of which were consistent with AMS-02 observations. 



  After these verifications, we tried to investigate CR anisotropy expected in the anisotropic diffusion model using these mothods.
  As an example, we adopted the Hundsdorfer-Verwer scheme, and performed a toy simulation in which the local grid of magnetic field direction was set to follow the IBEX experiment~\citep{2013ApJ...776...30F}, and found that the axis of dipole anisotropy would be consistent with the magnetic field direction.

  With more effort applied in the future, we would finally introduce a small enough grid size, and predict the CR anisotropy with these simulation methods and a realistic magnetic field.


In addition, the pseudo source method may provide an alternative measure when solving 3D nonlinear transportation equation, e.g., the nonlinear diffusive shock wave acceleration equation.

\acknowledgements

We thank Dr. M.-H. Cui for his helpful discussions, and Mr. Lu for editing the language of this paper. This work is supported by the National Key Research and Development Program of China (No. 2016YFA0400200, 2018YFA0404203) and Natural Sciences Foundation of China (11635011, 11851303, 11963004).


\vspace{5mm}
\software{GALPROP \citep{1998ApJ...509..212S} available at \url{https://galprop.stanford.edu}.}

\bibliographystyle{aasjournal}
\bibliography{ref}

\end{document}